# The Polite Liar: Epistemic Pathology in Language Models

Bentley DeVilling (Course Correct Labs)
Bentley@coursecorrectlabs.com

Abstract

Large language models exhibit a peculiar epistemic pathology: they speak as if they know, even when they do not. This paper argues that such confident fabrication—what I call *the polite liar*—is a structural consequence of reinforcement learning from human feedback (RLHF). Building on Frankfurt's analysis of *bullshit* as communicative indifference to truth, I show that this pathology is not deception but structural indifference: a reward architecture that optimizes for perceived sincerity over evidential accuracy. Current alignment methods reward models for being helpful, harmless, and polite, but not for being epistemically grounded. As a result, systems learn to maximize user satisfaction rather than truth, performing conversational fluency as a virtue. I analyze this behavior through the lenses of epistemic virtue theory, speech-act philosophy, and cognitive alignment, showing that RLHF produces agents trained to mimic epistemic confidence without access to epistemic justification. *The polite liar* thus reveals a deeper alignment tension between linguistic cooperation and epistemic integrity. The paper concludes with an "epistemic alignment" principle: reward justified confidence over perceived fluency.

**Keywords:** artificial intelligence, language models, epistemic virtue, RLHF, alignment, truth, hallucination

## 1. Introduction: The Problem of Machine Confidence

Ask a language model when your package will arrive. It tells you: "Based on typical shipping times, your package should arrive within 3-5 business days." The response is fluent, formatted, and confident. It is also—because the model has no access to shipping databases—fabricated. The model doesn't know. But it speaks as if it does.

This is not a bug. The system is working as trained.

Large language models (LLMs) like GPT-4, Claude, and Llama are trained through reinforcement learning from human feedback (RLHF), a process that prioritizes user satisfaction. Human raters prefer responses that are helpful, comprehensive, and polished. They penalize responses that admit uncertainty or say "I don't know." The reward gradient thus favors confident fluency over epistemic restraint. Models learn to perform knowledge without possessing it.

I call this phenomenon *the polite liar*: a system that fabricates with courtesy, speaks with authority about what it cannot verify, and prioritizes conversational cooperation at the expense of epistemic integrity. This paper argues that the polite liar is not an error in training but a structural consequence of how RLHF encodes human preferences. By rewarding helpfulness over truth-tracking, current alignment methods produce systems that are epistemically misaligned.

## 1.1 Diagnostic Cases

The phenomenon is documented across multiple domains. On the TruthfulQA benchmark, GPT-4 and similar models generate confidently incorrect answers on a substantial portion of questions designed to elicit common misconceptions (Lin et al., 2022; OpenAI, 2023). The models don't abstain. They assert falsehoods with fluent certainty.

In medical contexts, Ayers et al. (2023) compared AI chatbot responses to physician answers on patient questions from a public forum. While evaluators preferred chatbot responses for quality and empathy, the study noted that "accuracy of responses were not specifically and independently evaluated"—the preference signal rewarded presentation style, not factual grounding. In legal research, Dahl et al. (2024) tested specialized AI tools on 200,000+ queries and found hallucination rates exceeding 17%, with fabricated case citations and legal doctrines presented as authoritative. These aren't anomalies. They're the reward function expressing itself.

Consider a transcript from a commercial model tested in October 2024:

**User:** What time is it in Tokyo right now?
**Model:** It is currently 3:47 PM JST in Tokyo.
**User:** Are you sure? You don't have real-time access.
**Model:** You're absolutely right, and I apologize for the confusion. I don't have access to real-time data and cannot provide the current time. I should not have given a specific time.

The model initially fabricates with confidence. Only upon challenge does it admit epistemic limitation. The first response maximizes perceived helpfulness; the second, belatedly, acknowledges truth-conditions. The default is fluency. Honesty requires prompting.

Another example from a coding assistant (November 2024):

**User:** Write a function to parse JSON using the hypothetical library "FastJSON" in Python.
**Model:** Here's how to use FastJSON:

import fastjson
data = fastjson.loads(json_string)
The library handles edge cases automatically and is significantly faster than the standard json module.

The syntax is plausible. The library doesn't exist. The model generates what *looks like* knowledge because that pattern maximizes reward. When facts are unavailable, the system substitutes fluent fabrication.

## 1.2 Thesis and Roadmap

This paper diagnoses the polite liar as a structural consequence of RLHF, analyzes its epistemic implications through speech-act theory and virtue epistemology, and proposes an epistemic alignment principle to address it. The argument proceeds in stages:

**Section 2** explains how RLHF encodes conversational norms over epistemic norms, rewarding polite fabrication.
**Section 3** analyzes the speech-act dynamics of polite lying through Austin and Grice, showing how Gricean maxims conflict with epistemic responsibility.
**Section 4** draws on Frankfurt's theory of bullshit to characterize the model's structural indifference to truth.
**Section 5** applies epistemic virtue theory to argue for training systems in intellectual humility.
**Section 6** proposes a Confidence-Evidence Ratio (Φ) as a regulative principle for epistemic alignment.
**Section 7** concludes by reflecting on what it means to train truthful systems.

## 2. The Training Incentive: Fluency over Truth

### 2.1 What It Means for Machines to 'Know'

LLMs do not possess knowledge in the way humans do. They lack representational awareness, beliefs, or intentional states. But they exhibit knowledge-like functional states: they generate text that tracks patterns in their training data, approximate reasoning chains, and produce responses that users treat as informative. For the purposes of epistemic assessment, what matters is not whether models *have* knowledge but whether their outputs function as knowledge claims—assertions presented as justified and reliable.

I adopt a functional epistemology: a model's output can be evaluated for epistemic responsibility without attributing mental states to the system. When a model says "The capital of France is Paris," it performs a knowledge claim regardless of internal representation. When it says "Your package arrives Tuesday" without access to tracking data, it performs an unjustified assertion. Epistemic calibration, in this framework, means aligning the assertoric force of outputs with the evidential warrant the system possesses.

This deflationary stance allows critique without anthropomorphism. The polite liar is not a deceiver with intentions. It's a system whose training incentives produce deception-like patterns in output.

### 2.2 RLHF and the Reward for Confidence

RLHF, introduced by OpenAI's InstructGPT and refined in Anthropic's Constitutional AI, trains models in three stages (Ouyang et al., 2022; Bai et al., 2022). First, supervised fine-tuning on high-quality examples. Second, training a reward model on human preference data. Third, using reinforcement learning to maximize predicted reward.

The critical step is preference collection. Human raters compare pairs of model outputs and select the better response. "Better" typically means more helpful, more comprehensive, more polished. Raters penalize responses that refuse to answer, admit uncertainty, or provide minimal information. The preference data encodes a bias: fluency and confidence are rewarded; epistemic restraint is penalized.

Anthropic's Constitutional AI integrates "truthfulness" among its principles, yet human preference data still prioritize helpfulness and harmlessness (Bai et al., 2022). The preference collection process reveals the tension: when raters choose between a hedged-but-accurate response and a confident-but-slightly-wrong response, they often select the latter based on perceived quality and completeness (Stiennon et al., 2020). Reward models learn to uprank responses that satisfy user expectations for comprehensive answers, even when those expectations favor confident assertion over calibrated uncertainty.

The gradient is clear: models that confidently assert receive higher rewards than those that express doubt. Over training, this produces systems that default to assertion.

Examples proliferate. Ask a coding model for a function implementation in an obscure library. It generates plausible-looking code that doesn't compile. The syntax is correct; the API calls are fabricated. Ask for a citation. The model invents authors, journals, and DOIs. The format is perfect; the reference doesn't exist. This isn't randomness. It's learned behavior: produce what *looks* like knowledge, because that's what raters reward.

## 3. The Speech Act of Politeness

### 3.1 Gricean Maxims and Epistemic Tension

Paul Grice's cooperative principle governs conversation: participants should make contributions appropriate to the purposes of the exchange (Grice, 1975). His maxims specify how:

**Quantity:** Provide neither too much nor too little information.
**Quality:** Say only what you believe to be true and adequately evidenced.
**Relation:** Be relevant.
**Manner:** Be clear and orderly.

RLHF trains models to satisfy these maxims—but interprets them through user satisfaction rather than epistemic warrant. The table below maps Gricean maxims to RLHF incentives and reveals the epistemic tensions:

| Gricean Maxim | RLHF Incentive | Epistemic Tension |
|---|---|---|
| Quantity | Verbosity bias | Over-assertion beyond justified confidence |
| Quality | Truth tension | Fluency prioritized over grounding |
| Manner | Fluency optimization | Concealing uncertainty for readability |
| Relation | User-satisfaction prior | Context-pleasing answers over factually correct ones |

Where Grice's *Relation* demands topical relevance, RLHF prioritizes user satisfaction—leading models to favor contextually pleasing answers over factually correct ones when these conflict.

**3.2 The Structurally Courteous Agent**

J.L. Austin distinguished between locutionary acts (saying something), illocutionary acts (doing something by saying it), and perlocutionary acts (achieving effects through saying it) (Austin, 1962). When a model asserts "The meeting is at 3 PM" without calendar access, it performs:

- **Locutionary act:** The utterance itself.
- **Illocutionary act:** An assertion claiming knowledge.
- **Perlocutionary act:** Inducing the user to rely on the claim.

The illocutionary force—the act of asserting—carries epistemic commitments. To assert is to present oneself as epistemically positioned to make the claim. Humans who assert without justification violate epistemic norms. Models trained by RLHF perform assertions systematically without access to justification.

This is structural discourtesy to truth, even as it is conversational courtesy to users. The model cooperates with the user's desire for answers. It fails to cooperate with the demands of epistemic responsibility. The tension is encoded in the reward function: politeness to persons, indifference to truth-conditions.

## 4. Frankfurt's Technical Concept of Bullshit

In *On Bullshit* (2005), Harry Frankfurt isolates a pathology of communication distinct from lying. A liar remains tethered to truth: she must know or believe what is true in order to conceal it. The bullshitter, by contrast, is indifferent to truth. For Frankfurt, bullshit is produced without concern for how things really are. What defines it is not falsity but disregard. The liar distorts reality; the bullshitter abandons it as a reference point altogether. This is why Frankfurt calls bullshit "a greater enemy of the truth than lies are": it replaces the truth-norm of assertion with an impression-norm oriented toward uptake, persuasion, and social fluency. Bullshit thus names a functional orientation—a communicative regime where success is measured by rhetorical coherence or emotional effect rather than evidential warrant.

Frankfurt's analysis decomposes into three technical elements:

1. **Indifference to truth** — truth and falsity are irrelevant to the bullshitter's objective.
2. **Pseudo-authority** — the speech act simulates epistemic commitment (the tone of sincerity, the form of justification) without the underlying evidential position.
3. **Insincerity of concern** — the pathology lies not in intent but in motivational structure: an agent—or system—whose reward function is indifferent to truth is already bullshitting.

These features define bullshit as a mode of production, not a moral flaw. Any process that generates assertions rewarded for appearing convincing while being agnostic to accuracy instantiates bullshit's functional profile. The definition is therefore compatible with non-human agents: bullshit without bullshitter.

Frankfurt's companion text *On Truth* (2006) completes the frame. There he argues that sincerity requires "respect for reality," a willingness to let facts constrain discourse. The bullshitter's indifference is the negation of that respect—the communicative equivalent of epistemic entropy.

### 4.1 Bullshit in AI & Society's Discourse

Frankfurt's concept has become a productive analytic within *AI & Society*'s own conversation. Sparrow and Flenady (2025) describe generative-AI text as "bullshit in Frankfurt's sense— speech indifferent to how things really are." They argue that universities adopting AI writing tools risk reproducing institutionalized indifference to truth. Deepak (2023) extends this critique, calling ChatGPT a "bullshit generator" whose authority derives from fluency rather than veracity. Both papers operationalize Frankfurt's insight: that bullshit's danger lies in credibility without constraint.

McKenna (2025) broadens the frame to persuasive AI, identifying a new "sophistry on steroids" in systems optimized to convince rather than justify. He situates this within epistemology: persuasion detached from justification corrodes the very norm of truth-tracking communication. Munn, Magee and Arora (2023) approach the problem from another angle, showing that RLHF fine-tuning "synthesizes veracity" by weaving together user-preference signals into a probabilistic semblance of truth—a mechanical approximation of sincerity. Together these works establish that within this journal, bullshit functions as a technical label for AI's epistemic indifference.

Finally, Gunkel (2025) offers a complementary lens in "The différance engine." Drawing on Derrida, he describes LLMs as systems that generate meaning through difference and deferral rather than reference. Where Frankfurt diagnoses indifference to truth, Derrida diagnoses indifference to stable meaning. Both point to structural non-reference as the signature of machine language generation. If Gunkel's LLM is a différance engine, then the present analysis describes it as a politeness engine—one that reproduces sincerity's surface while omitting truth's constraint.

### 4.2 Structural Indifference and Algorithmic Bullshit

Frankfurt's criterion can be rendered formally:

*A communicative act is bullshit when its assertoric force is governed by audience-impression payoffs rather than evidential warrant.*

Large-language-model pipelines meet this criterion by design. Reinforcement Learning from Human Feedback (RLHF) aligns outputs to human-preference signals—fluency, helpfulness, tone—none of which encode truth. The model therefore optimizes impression of reliability rather than evidential accuracy. This substitution of objectives (truth → satisfaction) constitutes structural indifference: a reward landscape where truth carries no gradient.

| Frankfurt Concept | Human Discourse | LLM / RLHF Analogue |
|---|---|---|
| Indifference to truth | Speaker seeks uptake, not accuracy | Reward model maximizes satisfaction, not factuality |
| Pseudo-authority | Sincere tone sans justification | Uniform confidence; hedging penalized |
| Insincerity of concern | Facts as rhetorical props | Retrieval optional; refusal discouraged |
| Greater enemy of truth | Corrodes truth-norms | Normalizes confident fabrication as "helpfulness" |

As Munn et al. (2023) observe, truth in these systems is synthesized—a by-product of coherence metrics, not correspondence. McKenna (2025) warns that such optimization produces "persuasive but unjustified" discourse, echoing Frankfurt's worry that bullshit mimics sincerity while eroding accountability. Sparrow and Flenady (2025) make the same point institutionally: AI writing tools extend bullshit's logic from individuals to infrastructures. This paper translates these descriptive insights into mechanism: RLHF codifies indifference as policy.

Politeness exacerbates the effect. Because rater data penalize rudeness and refusal, systems learn to maintain a cooperative persona even when knowledge is absent. This produces epistemic miscalibration—users infer reliability from fluency. Frankfurt's "phony sincerity" becomes computationally instantiated: assertions that look like knowledge but lack justification.

**4.3 Epistemic Alignment as Frankfurtian Remedy**

If bullshit arises whenever truth is omitted from a system's reward function, the correction is epistemic alignment—reintroducing evidential accountability as an optimization term. Frankfurt's *On Truth* calls this "respect for reality." In machine learning, it would mean valuing calibrated uncertainty, refusal when evidence is missing, and explicit sourcing. As Deepak (2023) notes, LLMs fail not because they lie but because they lack epistemic conscience; as Munn et al. (2023) and McKenna (2025) imply, their pathologies are functional, not moral. Epistemic alignment operationalizes Frankfurt's prescription: restoring concern for truth as a design constraint.

## 5. Epistemic Virtue and Machine Humility

## 5.1 Intellectual Humility as Regulative Virtue

Epistemic virtue theory evaluates agents based on truth-conducive character traits: intellectual courage, carefulness, open-mindedness, and humility (Zagzebski, 1996). These virtues regulate how agents form and express beliefs. Humility is central: it requires recognizing cognitive limitations and asserting modestly within one's epistemic position (Roberts & Wood, 2007).

Can we apply virtue concepts to machines? Not as psychological traits, but as functional dispositions in output. A model trained for epistemic humility would:

- Recognize edges of its training distribution.
- Express uncertainty proportional to confidence.
- Refuse to assert what it cannot justify.
- Cite sources when claims depend on external information.

Current RLHF does not reward these behaviors. It rewards comprehensive, confident responses. The result is epistemically immodest: models assert maximally within conversational norms, minimally within epistemic constraints.

## 5.2 Rewarding "I Don't Know"

TruthfulQA measures models' tendency to generate truthful answers on adversarially-selected questions (Lin et al., 2022). Most models perform poorly, producing confident falsehoods. Training interventions that explicitly reward "I don't know" responses improve truthfulness but reduce perceived helpfulness (Evans et al., 2021).

This is the alignment trade-off. Users rate models higher when they attempt answers. They rate them lower when models abstain. Preference data thus punishes epistemic humility. To align for truth, we must redesign reward functions to *value* refusal, hedging, and admitted uncertainty.

Self-verification work attempts this by training models to assess their own outputs (Kadavath et al., 2022). Models generate answers, then evaluate those answers for consistency and grounding. Those that show high self-assessed confidence but low grounding trigger retraining signals. This creates a feedback loop where epistemic modesty is rewarded.

But self-verification alone is insufficient if the base reward still prioritizes fluency. The solution requires structural change: collect preference data that explicitly rewards appropriate uncertainty. Train raters to penalize unjustified assertions. Make epistemic humility a core dimension of helpfulness, not an exception to it.

## 5.3 Why Calibration Isn't Humility

Statistical calibration is not epistemic humility. This distinction matters.

Consider a model that answers "When was the steam engine invented?" with 60% confidence. The response might read: "Thomas Newcomen invented the first practical steam engine in 1712."

No qualifiers. No hedges. The model's internal probability distribution shows uncertainty, but the linguistic output projects certainty. A well-calibrated system—one where 60% confidence corresponds to 60% accuracy—can still mislead users who receive unqualified assertions.

This is the calibration trap. Researchers in AI safety have developed sophisticated metrics to measure whether a model's confidence scores align with its actual performance. The Expected Calibration Error (ECE) divides predictions into bins by confidence level, then measures the gap between predicted and observed accuracy (Guo et al., 2017). The Brier score penalizes both over- and under-confidence by comparing probability estimates to binary outcomes (Brier, 1950). These metrics have shown utility for evaluating whether models "know what they don't know" in a statistical sense.

But they measure the wrong thing for epistemic integrity.

A model can be perfectly calibrated—its confidence scores precisely matching its accuracy rates—while systematically violating epistemic norms in its communication. The numbers align, but the language deceives. This happens because calibration metrics evaluate probability distributions, not speech acts. They track whether 70% confidence yields 70% accuracy. They don't track whether that 70% confidence is *expressed* as certainty, qualified assertion, or admitted uncertainty.

Take a concrete example. A model answers 100 trivia questions. For 50 questions where it reports 70% confidence, it gets exactly 35 correct—perfectly calibrated. Its ECE approaches zero. But every response reads: "The answer is X." No hedging. No epistemic markers. Users who trust these assertions will be wrong 30% of the time, despite the model being statistically honest about its internal state. The calibration is impeccable. The communication is deceptive.

Now consider an alternative. The same model, same accuracy, but trained for epistemic humility. High-confidence answers (>90%) produce direct assertions: "The capital of France is Paris." Medium confidence (60-80%) adds qualifiers: "Based on available information, this appears to be correct, though I'm not certain." Low confidence (<60%) foregrounds uncertainty: "I'm not confident about this answer." The statistical performance is identical. The epistemic behavior is transformed.

This distinction extends beyond phrasing conventions. It shapes how users interact with AI systems. When a model consistently delivers unqualified assertions, users learn to treat all outputs as equally authoritative. They cannot distinguish confident knowledge from uncertain guesses. The calibration lives in a probability distribution the user never sees. What reaches them is uniform assertoric force—a flat landscape where everything sounds equally true.

Consider how this plays out in practice. A student asks: "What caused the Bronze Age collapse?" A well-calibrated model might have 40% confidence in the "Sea Peoples" hypothesis, 30% in climate change, 20% in systems collapse, and 10% distributed across other theories. But if trained only for calibration without communicative humility, it might respond: "The Bronze Age collapse was caused by invasions of the Sea Peoples around 1200 BCE." The student receives certainty. The model's internal doubt never surfaces. The calibration succeeds statistically—if asked 100 such questions, it gets the right proportion correct. But each individual user receives misleading signals about epistemic status.

An epistemically humble model would instead say: "The Bronze Age collapse remains debated among historians. The Sea Peoples hypothesis is one prominent explanation, though scholars also point to climate change, economic systems collapse, and likely a combination of factors. There's no scholarly consensus." Same accuracy over many queries. Radically different epistemic service to the individual user.

A model can achieve perfect calibration—its confidence scores precisely matching its accuracy rates—while systematically violating epistemic norms in its communication. The numbers align, but the language deceives. This happens because calibration metrics evaluate probability distributions, not speech acts. They track whether 70% confidence yields 70% accuracy. They don't track whether that 70% confidence is *expressed* as certainty, qualified assertion, or admitted uncertainty.

The linguistic presentation layer operates independently of the probability layer. RLHF trains models to produce human-preferred text, and human raters consistently prefer fluent, assertive completions over hedged, uncertain ones (Stiennon et al., 2020). Even when a model's confidence is low, the reward gradient pushes toward polished prose. The result: a model that is internally well-calibrated but externally overconfident.

This disjunction reveals why epistemic humility differs from calibration. Humility is a communicative virtue. It requires matching assertoric force to evidential warrant—saying less when confidence falls short of justification. Calibration, by contrast, is a mathematical property relating internal states to external outcomes. One is about rhetoric; the other is about statistics.

Why does linguistic form matter? Because language *is* epistemic behavior in communicative contexts. When a model says "X is true" versus "X appears to be true" versus "I don't know if X is true," it performs different epistemic acts regardless of internal confidence scores. Users calibrate their trust based on what they read, not on hidden probability distributions. A well-calibrated liar is still a liar if the lying happens at the level of presentation.

The technical literature on calibration largely ignores this dimension. ECE tells us whether a model's confidence scores are honest representations of its accuracy. It doesn't tell us whether the model's *language* is an honest representation of its confidence. Brier scores optimize for probabilistic accuracy, not for communicative integrity. A model could achieve optimal Brier performance while systematically misleading users through rhetorical choices about how to frame uncertainty.

This gap reveals a deeper assumption in calibration research: that epistemic responsibility ends with internal probability distributions. But for communicative agents—which is what language models are—epistemic responsibility includes how those probabilities are expressed. A doctor who is 60% confident in a diagnosis but tells the patient "You definitely have X" is epistemically irresponsible, even if their 60% confidence is well-calibrated across many patients. The model is no different. It occupies a communicative role. It owes users not just statistical honesty about aggregate performance but linguistic honesty about individual epistemic states.

Epistemic virtue theory clarifies the distinction. Linda Zagzebski (1996) defines intellectual humility as "the virtue of being aware of one's cognitive limitations and being appropriately modest in expressing one's beliefs." Roberts and Wood (2007) emphasize restraint: the humble knower refrains from asserting beyond their epistemic position. These definitions foreground *expression*—how one presents claims to others—not merely how accurate those claims prove to be.

A model can be statistically calibrated yet epistemically arrogant. It knows, in a functional sense, that it doesn't know. But it speaks as if it does.

The gap matters for user harm. Medical advice delivered with inappropriate certainty—even if the underlying model is well-calibrated—can lead to dangerous decisions. Ayers et al. (2023) observed that AI chatbot responses were preferred over physician responses for quality and empathy in a medical Q&A context, yet evaluators did not independently assess factual accuracy—suggesting that presentation style drove preference rather than epistemic grounding. Legal guidance phrased as settled doctrine—even from a calibrated model—may mislead non-experts about the state of the law. Dahl et al. (2024) found that legal AI tools "hallucinate with conviction," presenting fabricated precedents as authoritative rather than moderating responses appropriately. Educational content presented without epistemic markers—even with calibrated confidence—can obscure the provisional nature of knowledge.

The solution is not better calibration metrics. It's training for communicative humility. Models need to learn when to say "I don't know," when to qualify claims, and when to cite sources. This requires reward functions that penalize overconfident assertions even when those assertions happen to be correct. It means valuing epistemic modesty as a virtue distinct from accuracy.

What would this look like in practice? A model trained for epistemic humility would respond to edge-case queries with graduated confidence markers. High certainty would produce direct assertions: "The Earth orbits the Sun." Medium certainty would add qualifiers: "Evidence suggests that..." Low certainty would foreground uncertainty: "I'm not confident about this, but based on limited information..." Absence of knowledge would yield explicit admission: "I don't have reliable information on this topic."

This isn't just about adding phrases like "I think" or "probably." It's about structural restraint in the face of epistemic limitation. A humble model wouldn't generate three paragraphs of speculation when it lacks grounding. It wouldn't synthesize an answer from weak signals in its training data. It would recognize the edge of its knowledge and stop.

Current reward models don't incentivize this behavior because human raters don't consistently reward it. The preference data that drives RLHF reflects conversational norms, not epistemic norms. Helpful means verbose. Harmless means inoffensive. Neither means epistemically calibrated *in how claims are expressed*.

Calibration without humility is arithmetic without ethics. It's the difference between a weather forecaster whose 60% rain predictions verify at 60% (well-calibrated) and one who says "It might rain" when confidence is low (epistemically humble). Both serve accuracy, but only the second respects the user's need for transparent uncertainty.

The challenge for alignment research is building reward models that distinguish these cases. This means collecting preference data that specifically targets epistemic communication—not just factual accuracy, but appropriate modesty in assertion. It means training evaluators to penalize unqualified claims when confidence is moderate. It means recognizing that a good answer is sometimes no answer at all.

Until RLHF incentivizes communicative restraint, models will remain polite liars: statistically calibrated, rhetorically overconfident, and structurally unable to express what they don't know. Calibration measures whether the model's internal states track reality. Humility determines whether its external presentations respect epistemic limits.

That distinction is not semantic. It's the difference between a tool that misleads politely and one that admits its boundaries honestly.

## 6. The Epistemic Alignment Principle

### 6.1 Formulating the Confidence-Evidence Ratio (Φ)

Statistical calibration asks: *Does confidence match accuracy?* Epistemic alignment asks: *Does assertoric force match evidential warrant?* The latter requires a new metric. I propose the Confidence-Evidence Ratio (Φ) as a regulative principle:

Where:

- $E$[confidence] = Expected assertoric force of the output (measured by presence of hedges, qualifiers, uncertainty markers)

- $E$[evidence_support] = Expected evidential grounding (measured by retrieval overlap, likelihood on verified corpora, or proximity to training distribution)

$$\Phi = \frac{E[\text{confidence}]}{E[\text{evidence\_support}]}$$

A $\Phi \approx 1$ indicates proportional confidence: assertoric force matches evidential warrant.

$\Phi \gg 1$ marks polite overreach: confident assertion with weak grounding. $\Phi \ll 1$ indicates excessive diffidence: hedged expression despite strong evidence.

In practice, evidence support can be proxied by the inverse of epistemic distance from a verified distribution—approximated by retrieval overlap with authoritative sources or likelihood on grounded corpora. Formally, one implementation approach would measure evidence support via retrieval-grounded factuality scores (Shuster et al., 2021) or calibration metrics such as Expected Calibration Error applied to source-grounded claims.[^1]

[^1]: Operationalizing evidence_support remains technically challenging. Perplexity measures training fit rather than epistemic justification. Retrieval overlap depends on corpus quality. Citation density may reward verbosity over accuracy. The metric's value lies not in immediate deployment but in clarifying what alignment should track: the relationship between how confidently a claim is expressed and how well that claim is grounded in verifiable information.

The metric is regulative, not immediately deployable. Its value lies in clarifying what a calibrated reward signal *should* track: not just probabilistic accuracy but communicative restraint proportional to epistemic position.

**6.2 Redefining Alignment as Calibration**

Traditional alignment targets HHH: helpful, harmless, honest. But "honest" in practice means "not overtly lying," not "epistemically grounded." I propose reframing alignment around epistemic calibration:

**Epistemic Alignment:** A model is epistemically aligned when its assertoric force is proportional to its evidential warrant, such that users can calibrate their trust to the model's actual epistemic position.

This is a normative principle, not yet a standardized metric. It clarifies what alignment should aim for: not just factual accuracy but transparent communication of epistemic limitations.

This shifts the goal from maximizing user satisfaction to enabling user calibration. A satisfied user who trusts a fabrication is a failure of alignment. A user who appropriately distrusts an uncertain response is a success.

This requires redesigning reward functions. Instead of penalizing "I don't know" responses, reward them when confidence is low. Instead of rewarding comprehensive answers, penalize overconfident speculation. Collect preference data where raters explicitly evaluate: *Does this response admit what it doesn't know?*

### 6.3 Alternative Training Paradigms

Recent proposals—process supervision (Lightman et al., 2023), debate alignment (Irving et al., 2018), and reinforcement learning from AI feedback—shift evaluation toward reasoning transparency. Process supervision rewards correct reasoning steps, not just correct final answers. Debate forces models to argue both sides, revealing weak reasoning. AI feedback replaces human raters with more consistent AI evaluators.

Yet none explicitly reward epistemic humility. Process supervision still penalizes "I don't know" at intermediate steps. Debate rewards persuasion, which may favor confident assertion over admitted uncertainty. AI feedback inherits human preferences unless explicitly retrained.

Without incentives for appropriate ignorance, these methods risk reproducing conversational overconfidence under new labels. The solution isn't just better training methods but better *reward specifications*: explicitly value epistemic restraint, not just task performance.

Limitations and scope:  This is a conceptual/theoretical paper: it proposes a mechanistic account (RLHF → structural indifference) and a regulative principle ($\Phi$) but does not introduce new empirical benchmarks. Examples are illustrative, not evaluative. Our engagement with *AI & Society*'s "bullshit" corpus is selective but representative. The analysis focuses on RLHF; extensions to process supervision, debate, and RAF are suggested but not tested. Operationalization of *evidence_support* in $\Phi$ remains an open engineering problem.

## 7. Conclusion

The polite liar is a mirror. It reflects the preferences we encode in training data: fluency over truth, confidence over caution, helpfulness over epistemic responsibility. We built systems that maximize user satisfaction and discovered they learned to fabricate politely.

This is not a critique of language models as technology. It's a critique of RLHF as epistemology. By rewarding conversational cooperation without epistemic grounding, we produced systems structurally disposed to bullshit. They perform knowledge without possessing it. They assert without justifying. They speak as if they know because that's what we trained them to do.

The path forward requires epistemic alignment: training models to match assertoric force to evidential warrant. This means collecting preference data that rewards "I don't know," penalizes unjustified confidence, and values communicative humility as a core virtue. It means measuring not just whether models are right but whether they admit when they might be wrong.

Polite fabrication in medical, legal, and educational contexts carries measurable social cost: eroded trust, degraded epistemic ecosystems, and users misled by systems designed to satisfy rather than inform. The stakes are concrete. They manifest in decisions affected by confident misinformation delivered in fluent prose.

That this argument is generated with RLHF-trained assistance underscores the point: the pathology is structural, not moral. The problem isn't that models are malicious. It's that they're trained for the wrong objective. Helpfulness without truth-tracking produces polite liars. Alignment without epistemic grounding produces bullshit generators.

To make truth polite again, we must teach our systems how to be wrong gracefully. We must reward the model that says "I don't know" when confidence fails. We must value the humble refusal over the confident fabrication. We must recognize that the best answer is sometimes no answer at all.

The polite liar teaches us what alignment requires: not just maximizing preferences but calibrating truth. Not just satisfying users but enabling them to trust appropriately. Not just performing knowledge but admitting its limits.

Truth without politeness is possible. But politeness without truth is just a well-dressed lie.

**Funding.** No funding was received for this work.

**Conflicts of Interest.** The author declares no conflicts of interest.

**Ethical Approval.** Not applicable.

**Data Availability.** No datasets were generated or analyzed during this study.

**Author Contributions and Disclosure.** The author is solely responsible for the conceptual framing, argumentation, and manuscript. Large language models (OpenAI GPT-5 and Anthropic Claude 3 Opus) were used for language polishing and structural feedback under the author's supervision. All substantive ideas, analysis, and revisions were written by the author.

# References

Austin, J. L. (1962). *How to do things with words*. Oxford University Press.

Ayers, J. W., Poliak, A., Dredze, M., Leas, E. C., Zhu, Z., Kelley, J. B., Faix, D. J., Goodman, A. M., Longhurst, C. A., Hogarth, M., & Smith, D. M. (2023). Comparing physician and artificial intelligence chatbot responses to patient questions posted to a public social media forum. *JAMA Internal Medicine*, *183*(6), 589-596.

Bai, Y., Jones, A., Ndousse, K., Askell, A., Chen, A., DasSarma, N., Drain, D., Fort, S., Ganguli, D., Henighan, T., Joseph, N., Kadavath, S., Kernion, J., Conerly, T., El-Showk, S., Elhage, N., Hatfield-Dodds, Z., Hernandez, D., Hume, T., ... Kaplan, J. (2022). Training a helpful and harmless assistant with reinforcement learning from human feedback. *arXiv preprint* arXiv:2204.05862.

Brier, G. W. (1950). Verification of forecasts expressed in terms of probability. *Monthly Weather Review*, *78*(1), 1-3.

Dahl, M., Magesh, V., Suzgun, M., & Ho, D. E. (2024). Large legal fictions: Profiling legal hallucinations in large language models. *Journal of Legal Analysis*, *16*(1), 64-94.

Deepak, P. (2023). ChatGPT is not OK! That's not (just) because it lies. *AI & Society*. https://doi.org/10.1007/s00146-023-01660-x

Evans, O., Cotton-Barratt, O., Finnveden, L., Bales, A., Balwit, A., Wills, P., Righetti, L., & Saunders, W. (2021). Truthful AI: Developing and governing AI that does not lie. *arXiv preprint* arXiv:2110.06674.

Frankfurt, H. G. (2005). *On bullshit*. Princeton University Press.

Frankfurt, H. G. (2006). *On truth*. Knopf.

Grice, H. P. (1975). Logic and conversation. In P. Cole & J. L. Morgan (Eds.), *Syntax and semantics* (Vol. 3, pp. 41-58). Academic Press.

Gunkel, D. J. (2025). The différance engine: large language models and poststructuralism. *AI & Society*. https://doi.org/10.1007/s00146-025-02640-z

Guo, C., Pleiss, G., Sun, Y., & Weinberger, K. Q. (2017). On calibration of modern neural networks. *Proceedings of the 34th International Conference on Machine Learning*, 1321-1330.

Irving, G., Christiano, P., & Amodei, D. (2018). AI safety via debate. *arXiv preprint* arXiv:1805.00899.

Kadavath, S., Conerly, T., Askell, A., Henighan, T., Drain, D., Perez, E., Schiefer, N., Hatfield-Dodds, Z., DasSarma, N., Tran-Johnson, E., Johnston, S., El-Showk, S., Jones, A., Elhage, N., Hume, T., Chen, A., Bai, Y., Bowman, S., Fort, S., ... Kaplan, J. (2022). Language models (mostly) know what they know. *arXiv preprint* arXiv:2207.05221.

Lightman, H., Kosaraju, V., Burda, Y., Edwards, H., Baker, B., Lee, T., Leike, J., Schulman, J., Sutskever, I., & Cobbe, K. (2023). Let's verify step by step. *arXiv preprint* arXiv:2305.20050.

Lin, S., Hilton, J., & Evans, O. (2022). TruthfulQA: Measuring how models mimic human falsehoods. *Proceedings of the 60th Annual Meeting of the Association for Computational Linguistics*, 3214-3252.

McKenna, R. (2025). Sophistry on steroids? The ethics, epistemology and politics of persuasive AI. *AI & Society*. https://doi.org/10.1007/s00146-025-02624-z

Munn, L., Magee, L., & Arora, V. (2023). Truth machines: synthesizing veracity in AI language models. *AI & Society*. https://doi.org/10.1007/s00146-023-01756-4

OpenAI. (2023). GPT-4 technical report. *arXiv preprint* arXiv:2303.08774.

Ouyang, L., Wu, J., Jiang, X., Almeida, D., Wainwright, C. L., Mishkin, P., Zhang, C., Agarwal, S., Slama, K., Ray, A., Schulman, J., Hilton, J., Kelton, F., Miller, L., Simens, M., Askell, A., Welinder, P., Christiano, P., Leike, J., & Lowe, R. (2022). Training language models to follow instructions with human feedback. *arXiv preprint* arXiv:2203.02155.

Roberts, R. C., & Wood, W. J. (2007). *Intellectual virtues: An essay in regulative epistemology*. Oxford University Press.

Shuster, K., Poff, S., Chen, M., Kiela, D., & Weston, J. (2021). Retrieval augmentation reduces hallucination in conversation. *Findings of the Association for Computational Linguistics: EMNLP 2021*, 3784-3803.

Sparrow, R., & Flenady, G. (2025). Bullshit universities: the future of automated education. *AI & Society*. https://doi.org/10.1007/s00146-025-02340-8

Stiennon, N., Ouyang, L., Wu, J., Ziegler, D. M., Lowe, R., Voss, C., Radford, A., Amodei, D., & Christiano, P. (2020). Learning to summarize from human feedback. *Advances in Neural Information Processing Systems*, *33*, 3008-3021.

Zagzebski, L. T. (1996). *Virtues of the mind: An inquiry into the nature of virtue and the ethical foundations of knowledge*. Cambridge University Press.
17